\documentclass[12pt,twoside,a4paper]{article}
\usepackage{amsmath,amssymb,latexsym,theorem,natbib,epsfig,color,subfigure,footmisc,bbm}
\usepackage{multirow,graphicx,array,rotating,url}  
\usepackage{epstopdf,afterpage,wasysym}
\usepackage{verbatim}
\usepackage[normalem]{ulem}
\def\crps{\mathop{\hbox{\rm CRPS}}}
\def\crpss{\mathop{\hbox{\rm CRPSS}}}
\def\logs{\mathop{\hbox{\rm LogS}}}
\def\logss{\mathop{\hbox{\rm LogSS}}}

\numberwithin{equation}{section}

\title{Statistical post-processing of visibility ensemble forecasts}

\author{{S\'andor Baran}$^{1,*}$ and {M\'aria Lakatos}$^{1,2}$  \vspace*{0.5cm}\\
{\small $^1$Faculty of Informatics, University of Debrecen, Hungary}\\
{\small $^2$Doctoral School of Informatics, University of Debrecen, Hungary}
}  

\date{}

\begin{document}

\maketitle

\footnotetext[1]{Corresponding author: \url{baran.sandor@inf.unideb.hu}}
\begin{abstract}

 To be able to produce accurate and reliable predictions of visibility has crucial importance in aviation meteorology, as well as in water- and road transportation. Nowadays, several meteorological services provide ensemble forecasts of visibility; however, the skill, and reliability of visibility predictions are far reduced compared to other variables, such as temperature or wind speed. Hence, some form of calibration is strongly advised, which usually means estimation of the predictive distribution of the weather quantity at hand either by parametric or non-parametric approaches, including also machine learning-based techniques. As visibility observations -- according to the suggestion of the World Meteorological Organization -- are usually reported in discrete values, the predictive distribution for this particular variable is a discrete probability law, hence calibration can be reduced to a classification problem. Based on visibility ensemble forecasts of the European Centre for Medium-Range Weather Forecasts covering two slightly overlapping domains in Central and Western Europe and two different time periods, we investigate the predictive performance of locally, semi-locally and regionally trained proportional odds logistic regression (POLR) and multilayer perceptron (MLP) neural network classifiers. We show that while climatological forecasts outperform the raw ensemble by a wide margin, post-processing results in further substantial improvement in forecast skill and in general, POLR models are superior to their MLP counterparts.

\end{abstract}

\section{Introduction}
\label{sec1}
According to the definition of the World Meteorological Organization (WMO), visibility ``is the greatest distance at which a black object of suitable dimensions (located on the ground) can be seen and recognized when observed against the horizon sky'' \citep{wmo92}. This weather quantity plays a crucial role in aviation in all phases of flight and restricted visibility also makes both ship navigation and road transportation difficult. Hence, accurate and reliable visibility forecasts result in a direct economic benefit.

In general, weather forecasts are generated using numerical weather prediction (NWP) models, which are often run several times with varying model physics and/or initial conditions resulting in a probabilistic forecast represented by an ensemble of predictions \citep{btb15,b18a}. With the help of a forecast ensemble, besides providing point forecasts for a given location, time point, and forecast horizon, one can also assess forecast uncertainty and estimate probability distributions of future weather variables \citep{gr05}. However, in contrast e.g. to temperature, wind speed, or precipitation accumulation, most NWPs do not explicitly model visibility, so visibility forecasts must be derived from predictions of related quantities such as relative humidity or precipitation \citep{cr11}. Even nowadays only a few weather centres issue visibility ensemble forecasts. Examples include the multimodel Short-Range Ensemble Forecast System of the National Centers for Environmental Prediction covering the Continental US, Alaska, and Hawaii regions \citep{zdmqd09} or the Ensemble Prediction System (EPS) of the European Centre for Medium-Range Forecasts \citep[ECMWF;][]{mbpp96,ecmwf12}, where visibility is part of the Integrated Forecasting System (IFS) since 2015 \citep{ifs21}.

Despite the continuous improvement of the various operational EPSs over the past decades, ensemble forecasts still might display systematic bias or suffer from lack of calibration \citep[see e.g.][]{bhtp05}, that calls for some form of post-processing \citep{b18b}. Moreover, visibility forecasts are even more problematic, as their predictive performance, similar to total cloud cover \citep{ecmwfEval21}, is highly below the skill of ensemble forecast of e.g temperature, wind speed, pressure or precipitation accumulation \citep[see e.g.][]{zdgd12}.

Nowadays, one can select from a large collection of post-processing methods developed for a wide variety of weather variables; for an overview of the state-ot-the-art techniques see e.g. \citet{w18} or \citet{vbd21}. Parametric approaches like non-homogeneous regression \citep{grwg05} or Bayesian model averaging \citep{rgbp05} provide full predictive distribution of the investigated weather quantity, whereas quantile regression-based approaches \citep[see e.g.][]{fh07,brem19} result in probabilistic forecasts by estimating the quantiles of the forecast distribution. Moreover, in the last few years machine learning techniques such as quantile regression forests \citep{tmzn16}, distributional regression network \citep{rl18}, Berstein quantile network \citep{brem20}, or the classification and interpolation-based approach of \citet{sswh20} gain more and more popularity; for a recent comparison of neural network-based approaches we refer to \citet{sl22}.

While visibility forecasts can be considered as continuous (e.g. ECMWF ensemble forecasts are provided in 1 m steps), most synoptic observation (SYNOP) stations report observations according to the WMO suggestions, that is ``100 to 5 000 m in steps of 100 m, 6 to 30 km in steps of 1 km, and 35 to 70 km in steps of 5 km'' \citep[Section 9.1.2]{wmo18}, and observed visibility is rounded down to the nearest reported value. In this way, one has just 84 different values, so the corresponding predictive distribution is a discrete probability law. Hence, post-processing of visibility forecasts can be considered as an 84-group classification problem resulting in the probabilities of the different reported values. The situation is similar to the case of total cloud cover (TCC) reported in eighths of the sky covered by clouds called oktas taking just nine different values, only in this case the number of classes is almost ten times larger. To calibrate TCC ensemble forecasts, \citet{hhp16} proposed multiclass logistic regression \citep[MLR;][]{izen08} and proportional odds (or ordered) logistic regression \citep[POLR;][]{mccull80}, whereas \citet{bleab21} investigated the forecast skill of several machine learning-based classification methods such as multilayer perceptron neural network \citep[MLP;][]{dlbook}, gradient boosting machine \citep{fried01} and random forest algorithms \citep{breiman01}.

In the present work we investigate the predictive performance of POLR and MLP approaches to calibration of ECMWF visibility ensemble forecasts, as these two methods exhibit the best forecast skill among the techniques investigated by \citet{bleab21} in the context of post-processing ensemble predictions of TCC. As reference forecasts, we consider the raw visibility ensemble and climatology.

The paper is organized as follows. After a brief description of the studied visibility datasets in Section \ref{sec2}, we review the POLR and MLP methods in Section \ref{sec3} and also provide the investigated approaches to training data selection and the considered verification tools. Results of our two case studies are reported in Section \ref{sec4}, followed by short discussion and conclusions in Section \ref{sec5}.

\section{Data}
\label{sec2}

\begin{figure}[t]
   \centering
   \epsfig{file=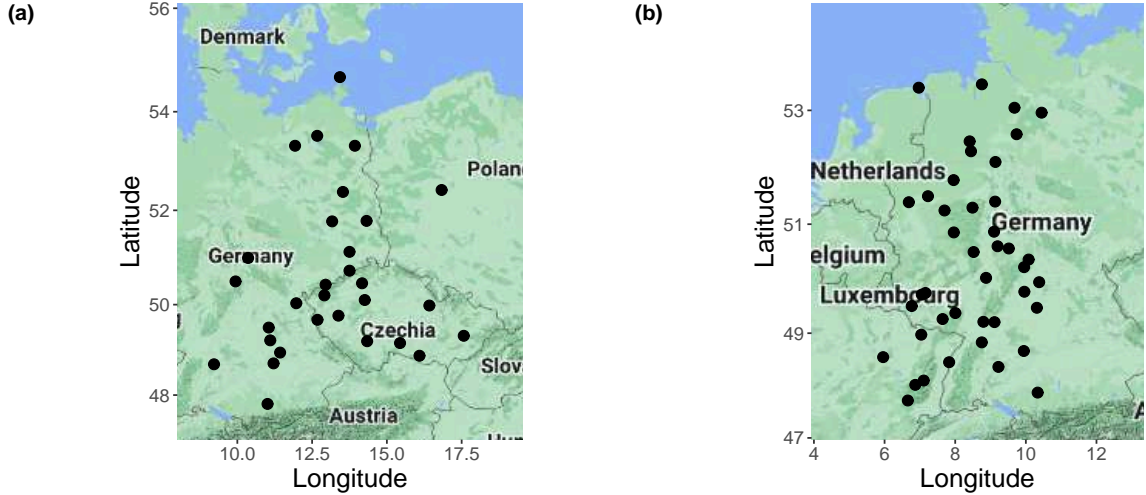, width=\textwidth} 
   \caption{Locations of SYNOP observation stations  corresponding to (a) ECMWF forecasts for 2020-2021; (b) EUPPBench benchmark dataset.}
   \label{fig:map}
 \end{figure}

We consider two datasets consisting of ECMWF visibility ensemble forecasts and corresponding
   validating observations covering different time periods but having slightly overlapping ensemble domains. As mentioned, observations are reported according to the WMO suggestions in values
\begin{equation*}
  {\mathcal Y}\!=\!\{0,100,200, \ldots, 4900, 5000, 6000, 7000, \ldots ,29000, 30000, 35000, 40000, \ldots , 65000, 70000\},
\end{equation*}
whereas the matching of forecasts (given in 1 m steps) and observations is performed by rounding down to the closest reported value.

In the case of the first dataset, as predictions we have the operational 51-member ECMWF visibility ensemble forecasts (control forecast (CTRL) and 50 members (ENS) generated using random perturbations) for calendar years 2020 and 2021 for 30 SYNOP stations in Germany, Czech Republic, and Poland (Figure \ref{fig:map}a). All forecasts are initialized at 0000 UTC and we consider 40 different lead times from 6 h to 240 h with a time step of 6 h. This data set is fairly complete as there are no missing observations at all and one has just two days when some predictions are missing; namely, there are no exchangeable ensemble members with forecast horizon 132 h initialized 2 June 2021 and 114 h initialized 17 December 2021.

We also investigate visibility data from the {\em EUPPBench\/} benchmark dataset \citep{eupp}, where besides the 51-member operational ECMWF ensemble the high-resolution (HRES) forecast is also available. The studied dataset consists of ensemble forecasts for calendar years 2017 -- 2018 initialized at 0000 UTC with a forecast horizon of 120 h and temporal resolution of 6 h for 42 SYNOP stations in Germany and France (Figure \ref{fig:map}b). Here around 1.5\,\% of the forecast cases is incomplete due to missing station observations.

\section{Post-processing methods and verification tools}
\label{sec3}

Following the notations of \citet{bleab21}, denote by \ $Y\in{\mathcal Y}=\{y_1,y_2, \ldots ,y_{84}\}$ \ the observed visibility for a given location and time point with \ $ {\mathcal Y}$ \ defined in Section \ref{sec2}, and let \
$\boldsymbol f=\big(f_1,f_2, \ldots ,f_{52}\big)$ \
be the corresponding ensemble forecast with a given forecast horizon, where \ $f_1=f_{HRES}$ \ denotes the high-resolution forecast, \ $f_2=f_{CTRL}$ \ is the control run and \ $f_3,f_4, \ldots ,f_{52}$ \ correspond to the 50 exchangeable members \ $f_{ENS,1},f_{ENS,2}, \ldots ,f_{ENS,50}$. \ As \ $Y$ \ is a discrete random variable, the predictive distribution of \ $Y$ \ is specified by conditional probabilities with respect to the ensemble forecast
\begin{equation}
  \label{eq:predDist}
  {\mathsf P}(Y=y_k \mid \boldsymbol f), \qquad k=1,2,\ldots ,84.
\end{equation}
Naturally, instead of the raw forecast \ $\boldsymbol f$, \  in \eqref{eq:predDist} the conditional probability can be taken with respect to any feature vector \ $\boldsymbol x$ \ derived from the ensemble and/or other covariates such as forecasts of additional weather quantities or station specific information like location, altitude or land use.

\subsection{Proportional odds logistic regression}
\label{subs3.1}

As mentioned in the Introduction, in the case of discrete weather quantities post-processing is reduced to a classification problem, where MLR is a frequently used and powerful tool. However, given an $M$-dimensional feature vector \ $\boldsymbol x$, \ an MLR model for an 84-group classification has \ $83(M+1)$ \ free parameters to be estimated from the training data. Hence, to avoid numerical issues in the estimation process, one requires an extremely large training data set consisting of past forecast-observation pairs.

A more parsimonious approach is the POLR model designed to fit ordered data, like the visibility observations at hand. In this case the conditional cumulative distribution of \ $Y$ \ with respect to an $M$-dimensional feature vector \ $\boldsymbol x$ \ is given as
\begin{equation}
  \label{eq:polr}
  {\mathsf P}\big(Y\leq y_k \mid \boldsymbol x\big)= \frac{{\mathrm e}^{\mathcal L_k(\boldsymbol x)}}{1+{\mathrm e}^{\mathcal L_k(\boldsymbol x)}},  \qquad \text{with} \qquad \mathcal L_k(\boldsymbol x):= \alpha_{k} + \boldsymbol x^{\top}\boldsymbol\beta, \quad \ k=1,2,\ldots, 84, 
\end{equation}
where \ $\alpha_k\in {\mathbb R}, \ \boldsymbol \beta \in {\mathbb R}^M$ \ and we assume \ $\alpha_{1}<\alpha_{2}< \cdots < \alpha_{84}$,  \ so the POLR model has \ $84+M$ \ unknown parameters.

\subsection{Multilayer perceptron neural network}
\label{subs3.2}
For both continuous and discrete weather quantities, the application of neural networks has increased in popularity for calibration. For the latter, instead of a parametric approach, one can use a classical feedforward multilayer perceptron (MLP) neural network for classification.  A classical MLP consists of multiple layers and neurons or nodes, each of which is a transformed (by an activation function) weighted sum of the node values from the previous layer plus an additional bias term. The features are "fed" to the input layer, and the predictions of the distribution of the various classes are made on the output layer. The number of hidden layers and neurons in them are tuning parameters of the network and provide the level of abstraction. For additional information, we refer to \citet{dlbook}.

\subsection{Training data selection}
\label{subs3.3}
Both parameters of the POLR model and weights of the MLP neural network are estimated with the help of training data, and the spatial and temporal composition of this set of forecast-observation pairs is a key issue in statistical post-processing. The basic approaches to spatial selection are local and regional (global) modelling \citep{tg10}. In the local case only past data of the station of interest are used to obtain the predictive distribution resulting in distinct POLR model parameters and MLP networks for the different stations.  Provided one has a long enough training period to avoid numerical issues \citep[for optimal training period lengths for different weather quantities see e.g.][]{hspbh14}, local models usually outperform their regional counterparts, which pool training data of the whole ensemble domain and all stations share the same set of POLR model parameters and a single MLP network. Regional modelling can be performed with the help of rather short training periods; however, this approach is not really suitable for large and heterogeneous domains. As a bridge between these two traditional spatial selection methods, \citet{lb17} propose a semi-local approach, where first clusters of stations with similar characteristics are formed using $k$-means clustering, then within each cluster a regional estimation is performed.

Concerning the temporal selection, a popular approach is the use of rolling training periods, where models are trained with the help of ensemble forecasts and validating observations from the preceding \ $n$ \ calendar days. This method enables models to quickly adapt e.g. to seasonal changes; however, in many situations the use of large fixed training data sets \citep[see e.g.][]{rl18,gzshf21} or yearly training considering data of the last couple of years before the year of the given date \citep[see e.g.][]{hhp16,bleab21} might also be beneficial.

\subsection{Verification scores}
\label{subs3.4}
As proposed by \citet{gbr07}, the evaluation of the predictive performance of probabilistic forecasts should be based on the idea of ``maximizing the sharpness of the predictive distributions subject to calibration''.  Calibration means a statistical consistency between forecasts and observations, while sharpness refers to the concentration of the predictive distribution.

A simple graphical tool for assessing calibration is the probability integral transform (PIT) histogram \citep[Section 9.5.4]{w19}. The PIT is defined as the value of the predictive cumulative distribution function (CDF) evaluated at the verifying observation, where for non-continuous laws randomization should be applied at points of discontinuity \citep{gr13}. For a properly calibrated probabilistic forecast the PIT follows a standard uniform distribution; moreover, the shape of the PIT histogram can provide hints to the possible reason for the lack of calibration.

The sharpness of a probabilistic forecast can be investigated by examining the widths of various prediction intervals. In the case studies of Section \ref{sec4}, similar to \citet{hhp16}, we compare the competing forecasts in terms of the average width of the centered 90\,\% prediction interval. Furthermore, one can also calculate the coverage of this interval defined as the proportion of validating observations located between the lower and upper 5\,\% quantiles of the predictive distribution. The deviation in coverage from the confidence level can serve as a measure of calibration.

Calibration and sharpness can also be assessed simultaneously by using proper scoring rules, which are loss functions assigning numerical values to forecast-observation pairs \ $(F,x)$. \ Here we consider the continuous ranked probability score \citep[CRPS;][Section 9.5.1]{w19} and the logarithmic score \citep[LogS;][]{good52}, being the most popular scoring rules in atmospheric sciences. Note, that in the case of visibility reported according to WMO suggestions (see Section \ref{sec2}), a forecast \ $F$ \ is characterized by a probability mass function (PMF) \ $p_F(y)$ \ specifying a discrete probability distribution on \ $\mathcal Y$.  \ Hence, the CRPS equals
\begin{equation}
  \label{eq:crps}
 \crps\big(F,x\big)=\sum_{k=1}^{84} p_F(y_k) \big| y_k -x \big| - \sum_{k=2}^{84} \sum_{\ell=1}^{k-1} p_F(y_k) p_F(y_{\ell}) \big| y_k - y_{\ell}\big|,
\end{equation}
whereas the LogS is defined as
\begin{equation*}
  \logs\big(F,x\big):= -\log \big(p_F(x)\big),
\end{equation*}
that is as the negative logarithm of the PMF at the observation.
Both CRPS and LogS are negatively oriented (the smaller the better), and expression \eqref{eq:crps} is the discrete form of the representation of the CRPS
\begin{equation*}
 \crps\big(F,x\big)={\mathsf E}|X-x|-\frac 12 {\mathsf E}|X-X'|
\end{equation*}
provided by \citet{gr07}, where \ $X$ \ and \ $X'$ \ are independent random variables with distribution \ $F$ \ and finite first moment.

For each of the investigated forecast horizons, the predictive performance of various probabilistic forecasts is quantified by the mean CRPS \ ($\overline\crps$) \ and mean LogS \ ($\overline\logs$) \ over all forecast cases in the verification data. Furthermore, one can obtain a deeper insight into the smaller differences in forecast skill of the compering predictions by examining continuous ranked probability skill scores (CRPSS) and logarithmic skill scores (LogSS), which measure the improvement in terms of CRPS and LogS of a forecast \ $F$ \  with respect to a reference forecast \ $F_{ref}$, \ respectively \citep[see e.g.][]{murphy73,gr07}. CRPSS and LogSS are defined as
\begin{equation*}
 \crpss := 1- \frac{\overline{\crps}}{\overline{\crps}_{ref}} \qquad \text{and} \qquad \logss := 1- \frac{\overline{\logs}}{\overline{\logs}_{ref}},
\end{equation*}
where \ $\overline\crps, \ \overline\logs$ \ and \  ${\overline\crps}_{ref}, \ {\overline\logs}_{ref}$ \ denote the mean score values corresponding to forecasts \ $F$ \ and \ $F_{ref}$, \ respectively. Note that skill scores are positively oriented, so larger values mean better predictive performance.

Furthermore, as point forecasts, we consider the ensemble mean and the means of the predictive distributions of the competing forecasts, which are evaluated with the help of the corresponding root mean squared errors (RMSEs).

Finally, the uncertainty of the verification scores and statistical significance of the score differences are addressed by providing confidence intervals for the skill scores. The standard deviations required for the confidence bounds are calculated from 2000 block bootstrap samples obtained using the stationary bootstrap scheme, where the mean block length is computed according to \citet{pr94}.

\section{Results}
\label{sec4}

In the case studies of Sections  \ref{subs4.2} and \ref{subs4.3}, the forecast skill of the POLR and MLP approaches described in Sections \ref{subs3.1} and \ref{subs3.2}, respectively, are evaluated.  Both methods require training data, which should be large enough to allow reliable modelling. We consider 350-day rolling training periods, which in the case of regional estimation means 10500 forecast cases for the first dataset and 14700 forecast cases for the second one for each training step. In what follows, regional POLR and MLP models are referred to as {\em POLR-R} and {\em MLP-R}, respectively. We also investigate clustering-based semi-local modelling, where the clusters of stations are derived using $k$-means clustering of feature vectors depending on station climatology over the training period. In particular, each station is represented by a three-dimensional feature vector providing the frequencies of visibility intervals 0 to 5000 m, 5000 to 30000 m and 30000 to 70000 m. With the shift of the training data the clusters are recalculated dynamically and in general, 4 clusters are formed in the case of forecasts for 2020 -- 2021 and 8 in the case of the EUPPBench data, provided each cluster contains at least 4 locations. Otherwise, the number of clusters is reduced and the clustering-based estimation might even fall back to regional modelling. For the semi-local POLR and MLP approaches notations {\em POLR-C} and {\em MLP-C} are used and note that the above feature vectors and the applied number of clusters are based on a preliminary analysis of the climatology of each station.  Finally, local training is also investigated, the corresponding models are denoted as {\em POLR-L} and {\em MLP-L}.

In both case studies the forecast skill of the competing post-processing methods is tested on data for a complete calendar year. As reference forecasts we consider the raw ECMWF visibility ensemble and climatology, where observations of a rolling training period of a given length are considered as a forecast ensemble.

Before calculating the input features of the POLR and MLP models, a normalization of the forecasts is performed; namely, all forecast values are divided by 70000. Then we consider the following set of predictors: the (normalized) control member of the ensemble \ $\tilde f_{CTRL}:=f_{CTRL}/70000$, \ the mean \ $\overline f_{ENS}$ \ of the 50 (normalized) exchangeable members, the variance \ $s^2$ \ of the 51-member (normalized) operational ensemble, in the case of the EUPPBench dataset the (normalized) high-resolution forecast \ $\tilde f_{HRES}:=f_{HRES}/70000$, \ and the proportions \ $p_1, \ p_2$ \ and \ $p_3$ \ of ensemble members predicting visibility up to 1000 m, 1000 to 2000 m and more than 30000 m, respectively. Note that normalization is required to have all input features commensurate. Finally, following the suggestions of e.g. \citet{dmmz17}, seasonal variations are addressed by adding annual base functions
  \begin{equation*}
    \beta_1(d):=\sin \big(2\pi d/365 \big) \qquad \text{and} \qquad  \beta_2(d):=\cos \big(2\pi d/365 \big)
  \end{equation*}
to the input features, where \ $d$ \ denotes the day of the year.

\subsection{Implementation details}
\label{subs4.1}

POLR models are fit using the {\tt R} package {\tt MASS} \citep{vr02}, while
MLP classification is based on the package {\tt RSNNS} making the Stuttgart Neural Network Simulator \citep{snns94} available in {\tt R}. The neural network has two hidden layers with 25 -- 25 neurons, the maximal number of iterations to learn is restricted to 200, the learning rate is $0.2$ and both hidden layers use the logistic activation function.

To handle numerical problems with LogS calculation caused by zero predicted probabilities, we follow the procedure applied by \citet{hhp16} and \citet{bleab21}. Extremely low values of the PMF are replaced with a probability \ $p_{\min}$, \ which ensures that the corresponding reported visibility at the given observation hour appears at least once a year with probability \ $\pi$. \ This means that for an observation \ $y_j$ \ instead of \ $p_F(y_j)$ \ we consider \ $\max\big\{p_{\min},p_F(y_j)\big\}$, \ where \ $p_{\min}=1-(1-\pi)^{1/365}$,  \ and normalize the obtained values to get a PMF again. For \ $\pi = 0.01$ \ this approach results in  \ $p_{\min}=2.75\times 10^{-5}$.

\subsection{Calibration of 51-member visibility ensemble forecasts}
\label{subs4.2}

\begin{figure}[t]
\begin{center}
\epsfig{file=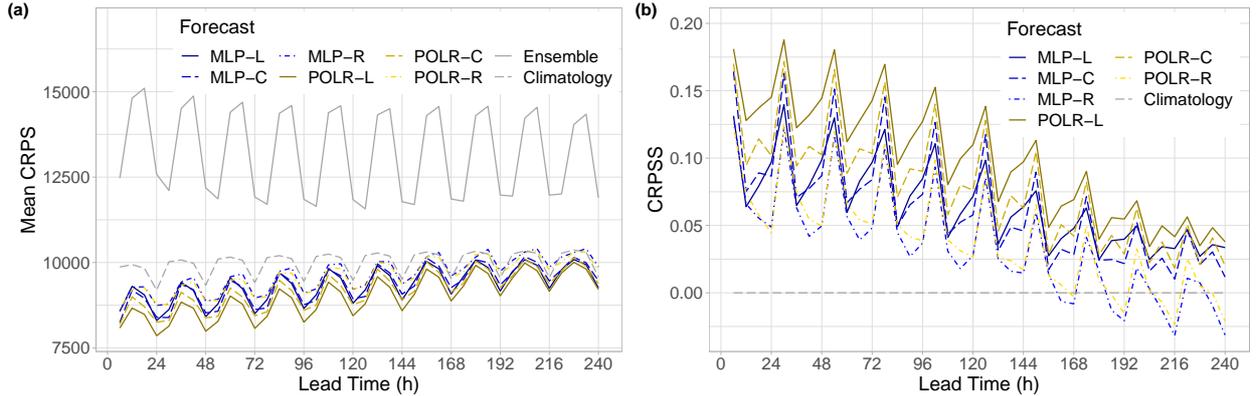, width=\textwidth}
\end{center}
\caption{Mean CRPS of post-processed, raw and climatological visibility forecasts for calendar year 2021 (a) and CRPSS of post-processed forecasts with respect to climatology (b) as functions of the lead time.}
\label{fig:crps_crpss}
\end{figure}

\begin{table}[t]
  \begin{center}
    \begin{tabular}{l|c|c|c|c|c|c|c}
  &MLP-L&MLP-C&MLP-R&POLR-L&POLR-C&POLR-R&Climatology\\ \hline
  CRPS&70.65\,\%&70.75\,\%&72.94\,\%&68.09\,\%&69.60\,\%&72.52\,\%&75.60\,\% \\
LogS&45.31\,\%&44.10\,\%&44.88\,\%&44.28\,\%&43.45\,\%&44.44\,\%&66.05\,\%                         
\end{tabular}
\end{center}
\caption{Overall mean CRPS/LogS of post-processed and climatological visibility forecasts for calendar year 2021 as proportion of the mean CRPS/LogS of the raw ECMWF ensemble.}
\label{tab1}
\end{table}

We study post-processing of visibility ensemble forecasts with the help of locally, semi-locally and regionally trained POLR and MLP models based on the 8-dimensional feature vector \ $\big(\tilde f_{CTRL},\overline f_{ENS},s^2,p_1,p_2,p_3,\beta_1,\beta_2\big)^{\top}$. \ Following the suggestions of \citet{hhp16}, for the POLR models the weights of \ $\tilde f_{CTRL}$,  \ and \  $\overline f_{ENS}$ \ are kept non-negative by excluding iteratively covariates with negative coefficients. The predictive performance of the competing post-processing methods is tested on data for the calendar year 2021. In the case of the reference climatological forecasts we consider a 30-day rolling training period, which value is chosen by the comparison of the overall mean CRPS, RMSE of mean forecasts and coverage of 90\,\% central prediction intervals for training periods of length 25, 30, \ldots , 50 and 350 days.

\begin{figure}[t]
\begin{center}
\epsfig{file=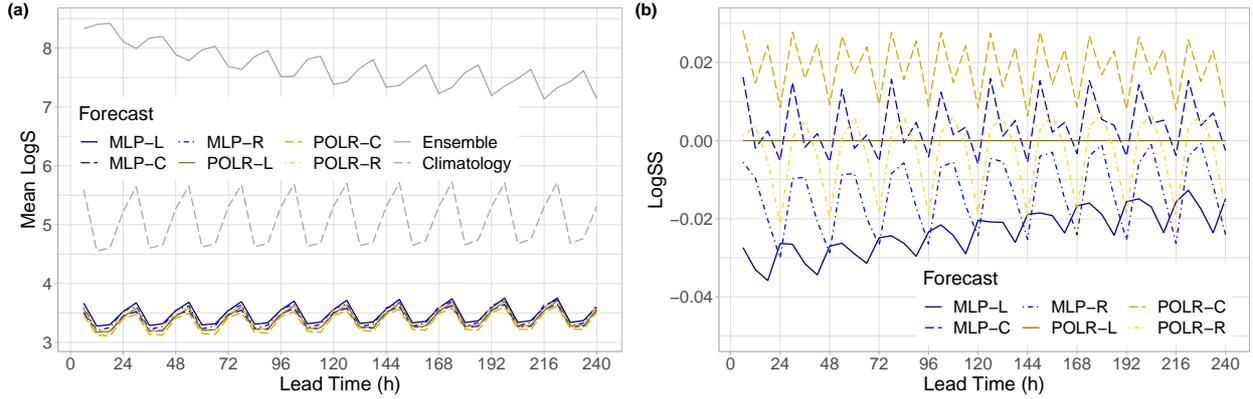, width=\textwidth}
\end{center}
\caption{Mean LogS of post-processed, raw and climatological visibility forecasts for calendar year 2021 (a) and LogSS of post-processed forecasts with respect to the POLR-L model (b) as functions of the lead time.}
\label{fig:logs_logss}
\end{figure}

\begin{figure}[t]
\begin{center}
\epsfig{file=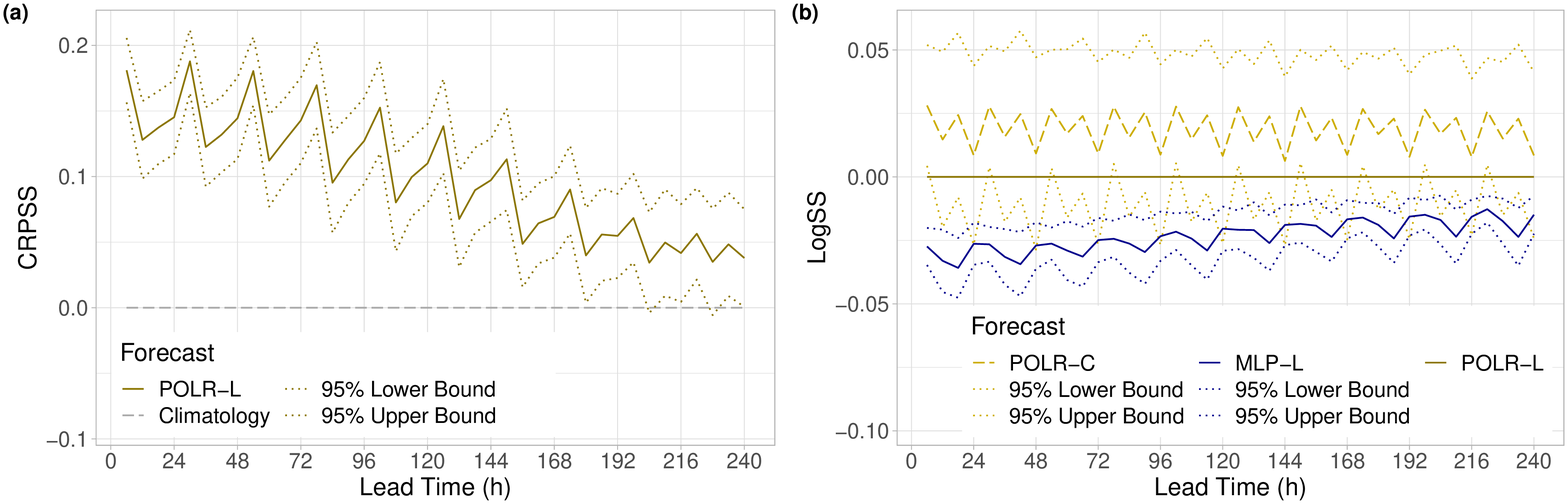, width=\textwidth}
\end{center}
\caption{CRPSS of POLR-L post-processed visibility forecasts for calendar year 2021 with respect to climatology (a) and LogSS of POLR-C and MLP-L approaches with respect to the POLR-L model (b)  together with 95\,\% confidence intervals as functions of the lead time.}
\label{fig:crpss_logss}
\end{figure}

\begin{figure}[h!]
\begin{center}
  \epsfig{file=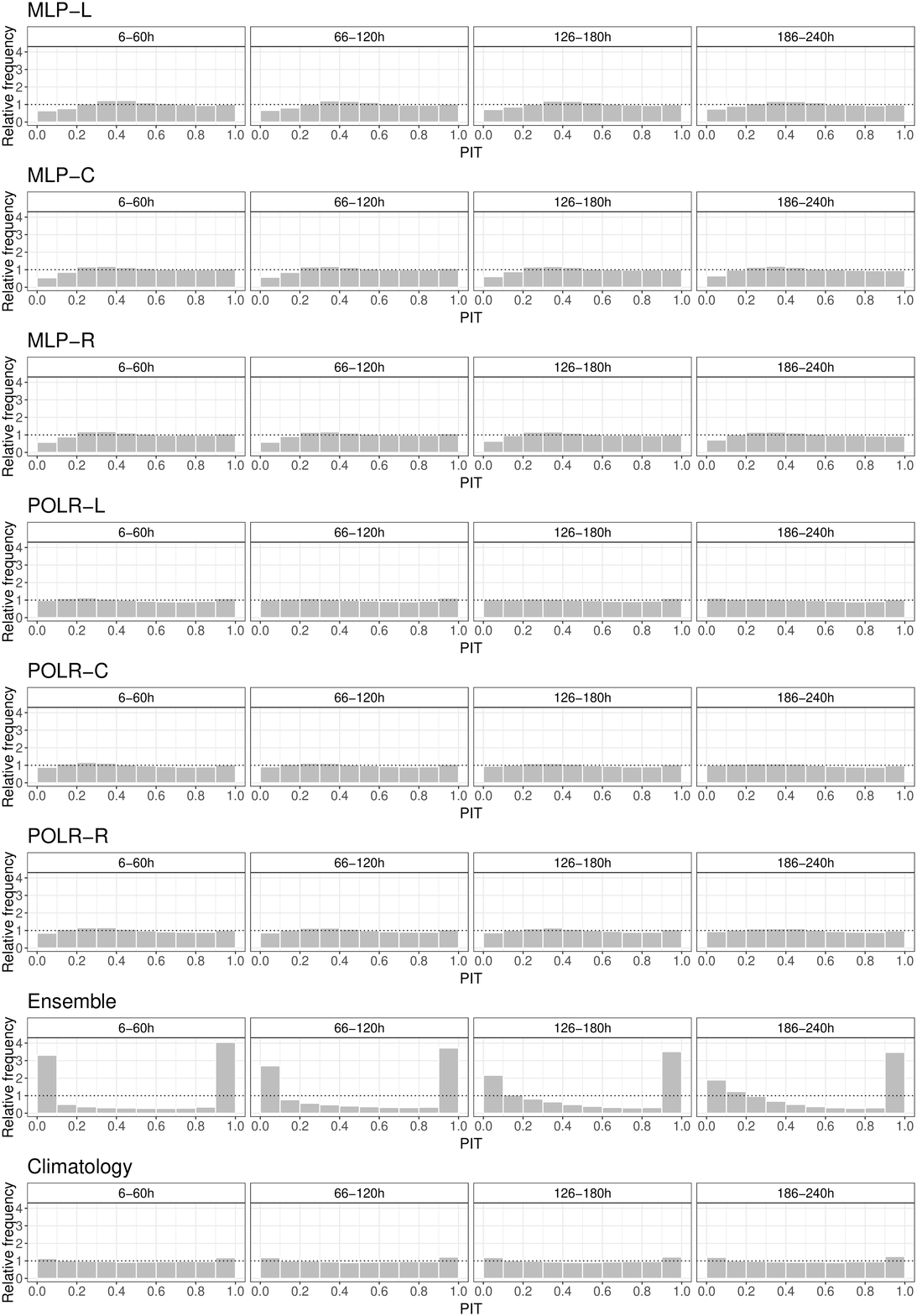, width=.9\textwidth}
\end{center}
\caption{ PIT histograms of post-processed, climatological and raw visibility forecasts for calendar year 2021 for lead times 6–60 h, 66–120 h, 126–180 h and 186–240
h.}
\label{fig:pit}
\end{figure}

According to Figure \ref{fig:crps_crpss}a, in terms of the mean CRPS all post-processing methods outperform the raw ECMWF ensemble forecast for all forecast horizons by a wide margin and climatology also performs rather well. The clear diurnal cycle in the mean CRPS corresponds to the different observation times (0000, 0600, 1200 and 1800 UTC). Note that the superiority of climatology over the raw visibility forecasts is fully in line with the results of \citet{cr11}. A deeper insight into the differences between the various calibration techniques can be obtained from Figure \ref{fig:crps_crpss}b depicting the CRPSS values with respect to climatology.
In general, clustering-based modelling is superior to regional, and the advantage of post-processing with respect to climatology decreases with the increase of the lead time. For POLR models the chosen training period length is long enough even for reliable local estimation and the POLR-L approach results in the highest skill score for all lead times. This is not the case for the MLP models, where the CRPSS of the semi-locally estimated MLP-C is often above the skill score of the MLP-L, especially for lead times corresponding to 0600 UTC. POLR models outperform their MLP counterparts for most of the forecast horizons; however, up to 156 h even the poorest MPL-R has better skill than climatology.  A possible ranking of the competing methods can be obtained from Table \ref{tab1} giving the overall mean score values of calibrated and climatological forecasts as proportions of the corresponding mean score of the raw ECMWF ensemble. POLR-L results in the lowest mean CRPS, closely followed by POLR-C. 

As Figure \ref{fig:logs_logss}a and Table \ref{tab1} show, the difference between raw and post-processed forecasts in terms of the mean LogS is even more pronounced than in terms of the mean CRPS; however, the general picture slightly changes. On the one hand, all post-processing methods outperform climatology by a wide margin. On the other hand, in terms of the LogS, clustering-based modelling is clearly superior to local estimation. According to Figure \ref{fig:logs_logss}b, for most of the forecast horizons MLP-L underperforms even MLP-R, and at lead times corresponding to 0600 and 1200 UTC, POLR-R also results in positive LogSS with respect to POLR-L.

To enlighten the uncertainty of the investigated verification scores and statistical significance of the score differences, in Figure \ref{fig:crpss_logss} the CRPSS and LogSS values of some post-processing approaches are accompanied with 95\,\% confidence intervals. According to Figure \ref{fig:crpss_logss}a, the advantage of the best-performing POLR-L model over climatology in terms of the mean CRPS is clearly significant for all lead times but 204 h and 228 h. Regarding the LogS (Figure \ref{fig:crpss_logss}b), POLR-L significantly outperforms MLP-L for all lead times, whereas the difference between POLR-L and the best performing POLR-C is significant at a 5\,\% level only at forecast horizons corresponding to 0600 UTC.

\begin{table}[t]
  \begin{center}
    \begin{tabular}{c|c|c|c|c|c}
MLP-L&MLP-C&MLP-R&POLR-L&POLR-C&POLR-R\\ \hline
$1.98\times 10^{-15}$&$6.54\times 10^{-14}$&$1.15\times 10^{-7}$&$4.31\times 10^{-5}$&$1.54\times 10^{-4}$&$2.00\times 10^{-5}$ \end{tabular}
\end{center}
\caption{Overall mean $p$-values of the $\alpha^0_{1234}$ tests for uniformity of the PIT values for calendar year 2021.}
\label{tab2}
\end{table}

Furthermore, PIT histograms of Figure \ref{fig:pit} also demonstrate the positive effect of post-processing and the superiority of climatology over the raw visibility ensemble forecasts. The histograms of the latter are rather U-shaped with a more and more pronounced bias with the increase of the forecast horizon. All other forecasts result in PIT values that are much closer to the desired standard uniform distribution; however, climatology still exhibits a small underdispersion, whereas MLP post-processed predictions are slightly overdispersive. For all calibrated forecasts the moment-based $\alpha^0_{1234}$ test \citep{k15} rejects uniformity at a 5\,\% level of significance for all 40 investigates lead times. Nevertheless, a possible ranking of the competing post-processing approaches in terms of goodness of fit of PIT can be obtained from Table \ref{tab2} reporting the mean $p$-values of the $\alpha^0_{1234}$ tests over all considered lead times (the larger the better). Note that for raw and climatological forecasts the PIT values are concentrated around 52 and 31 bins (ensemble size + 1), respectively. Hence, these PIT histograms fall back to the corresponding verification rank histograms \citep[][Section 9.7.1]{w19}. 

\begin{figure}[t]
\begin{center}
\epsfig{file=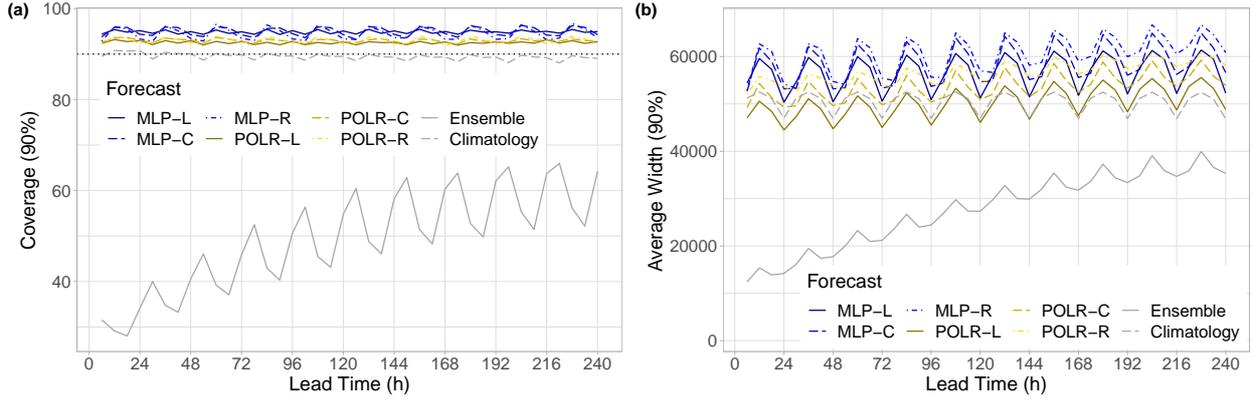, width=\textwidth}
\end{center}
\caption{Coverage (a) and average widths (b) of 90\,\% central prediction intervals of raw and post-processed visibility forecasts for calendar year 2021 as functions of the lead time. In panel (a) the ideal coverage is indicated by the horizontal dotted line.}
\label{fig:cov_aw}
\end{figure}

The improved calibration of post-processed and climatological forecasts can also be observed in Figure \ref{fig:cov_aw}a showing the coverage of 90\,\% central prediction intervals. For the raw forecasts this coverage ranges from 28\,\% to 66\,\% and shows an increasing trend with a clear diurnal cycle. For post-processed and climatological forecasts the diurnal cycle is far less pronounced and one cannot observe dependence on the forecast horizon. The coverage of climatology is almost perfect, whereas post-processed forecasts result in coverages slightly above 90\,\%. For the POLR models the mean absolute deviations over all lead times from the desired 90\,\% are 2.53\,\% (POLR-L),  2.96\,\% (POLR-C) and 2.99\,\% (POLR-R), while for the MLP approaches one gets 4.86\,\% (MLP-R),  4.98\,\% (MLP-C) and  4.52\,\% (MLP-R). Note that for climatological forecasts this mean absolute deviation is just 0.69\,\%. Naturally, the price for the better calibration of post-processed and climatological forecasts has to be paid in the loss of the sharpness.  As depicted in Figure \ref{fig:cov_aw}b, the raw ECMWF visibility ensemble results in far the narrowest 90\,\% central prediction intervals and in general, the average width values of the different forecasts are fairly consistent with the corresponding coverages.

\begin{figure}[t]
\begin{center}
\epsfig{file=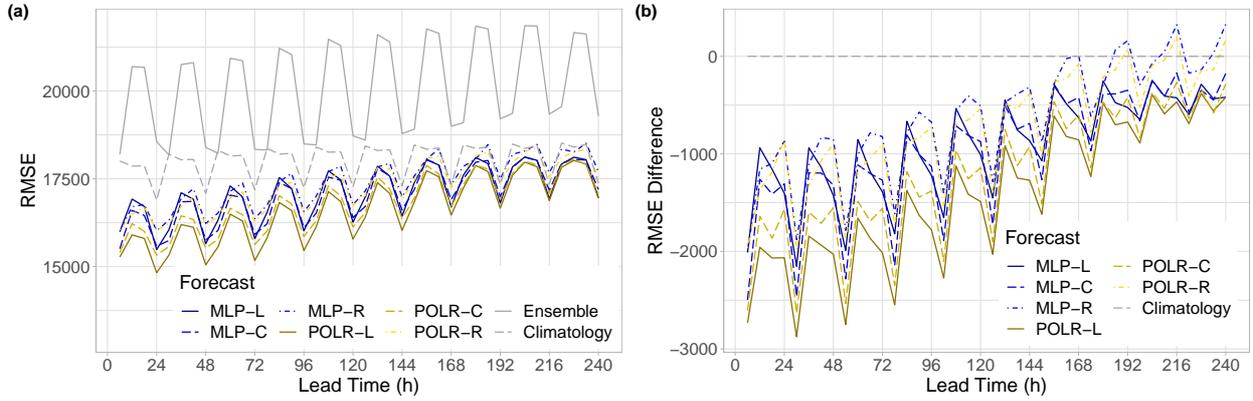, width=\textwidth}
\end{center}
\caption{RMSE of the mean forecasts for calendar year 2021 (a) and difference in RMSE from climatology (b) as functions of the lead time.}
\label{fig:rmse_rmsed}
\end{figure}

Finally, according to  Figure \ref{fig:rmse_rmsed}a, compared with the raw ensemble all POLR and MLP models substantially improve the accuracy of the mean forecast, whereas the advantage of climatology is far less pronounced, especially for shorter forecast horizons. Note that the ranking of the different forecasts with respect to the RMSE of the mean  (Figure \ref{fig:rmse_rmsed}b) is completely in line with the ordering based on the mean CRPS (Figure \ref{fig:crps_crpss}b), and the POLR-L approach again outperforms the competitors for all lead times.

Besides the 30-day reference climatology, the predictive performance of climatological forecasts based on a 350-day rolling training period was also tested. However, such a long training period is unlikely to capture the seasonal variations in visibility. In terms of the mean CRPS and RMSE of the mean, this forecast is far behind the various MLP and POLR models, its mean absolute deviation in 90\,\% coverage from the required value is 3.90\,\%, combined with an average width similar to the MLP approaches. The only score where the performance of the 350-day climatology is reasonable is the LogS, at 192 h, 216 h and 240 h it can compete with the regional versions of both studied post-processing techniques.

\subsection{Calibration of EUPPBench visibility ensemble forecasts}
\label{subs4.3}

\begin{figure}[t]
\begin{center}
\epsfig{file=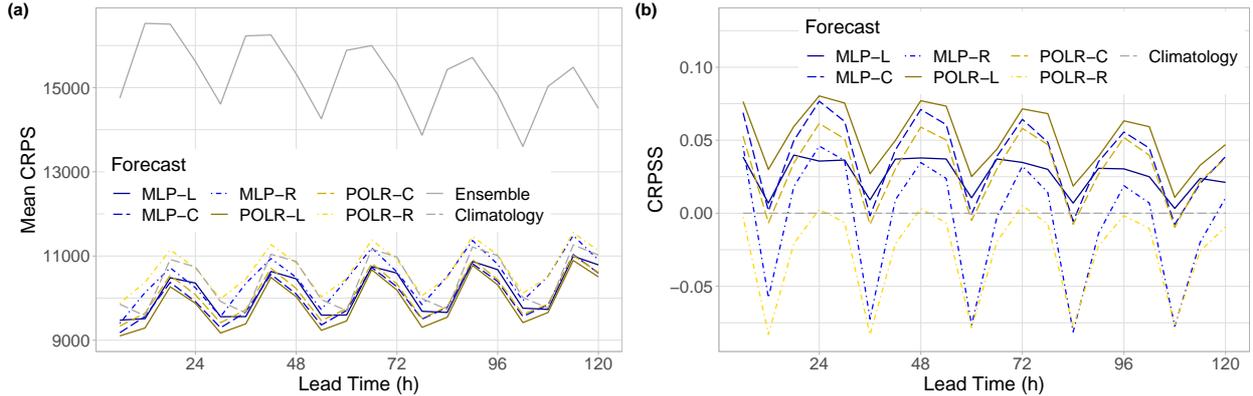, width=\textwidth}
\end{center}
\caption{Mean CRPS of post-processed, raw and climatological EUPPBench visibility forecasts for calendar year 2018 (a) and CRPSS of post-processed forecasts with respect to climatology (b) as functions of the lead time.}
\label{fig:crps_crpssB}
\end{figure}

\begin{table}[t]
  \begin{center}
    \begin{tabular}{l|c|c|c|c|c|c|c}
  &MLP-L&MLP-C&MLP-R&POLR-L&POLR-C&POLR-R&Climatology\\ \hline
  CRPS&66.35\,\%&65.53\,\%&68.47\,\%&64.67\,\%&66.05\,\%&69.96\,\%&68.19\,\% \\
LogS&45.48\,\%&43.91\,\%&44.45\,\%&44.86\,\%&43.66\,\%&44.55\,\%&64.39\,\%         \end{tabular}
\end{center}
\caption{Overall mean CRPS/LogS of post-processed and climatological EUPPBench visibility forecasts for calendar year 2018 as proportion of the mean CRPS/LogS of the raw ECMWF ensemble.}
\label{tab3}
\end{table}

As mentioned, in post-processing of the 52-member EUPPBench benchmark visibility forecasts as an additional predictor we consider the normalized high-resolution forecast and investigate the performance of POLR and MLP models with an extended feature vector  \
$$\big(\tilde f_{HRES},\tilde f_{CTRL},\overline f_{ENS},s^2,p_1,p_2,p_3,\beta_1,\beta_2\big)^{\top}.$$
Similar to the previous case study, for the local, semi-local and regional POLR models the coefficients of \ $\tilde f_{HRES}, \ \tilde f_{CTRL}$  \ and \ $\overline f_{ENS}$ \ are forced to be non-negative. Furthermore, the performance of the various forecasts is compared using forecast-observation pairs for the calendar year 2018 and following the same selection procedure as in Section \ref{subs4.2}, for the reference climatology now a 40-day training period is chosen. Finally, we tested again the skill of those climatological forecasts, that are based on the same 350-day training period as the post-processing approaches.

\begin{figure}[t]
\begin{center}
\epsfig{file=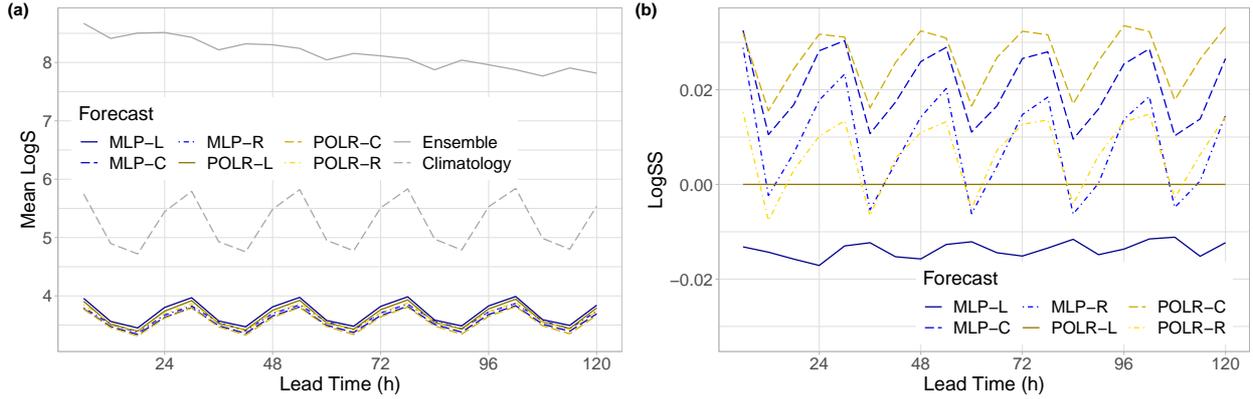, width=\textwidth}
\end{center}
\caption{Mean LogS of post-processed, raw and climatological EUPPBench visibility forecasts for calendar year 2018 (a) and LogSS of post-processed forecasts with respect to climatology (b) as functions of the lead time.}
\label{fig:logs_logssB}
\end{figure}

In line with the case study of Section \ref{subs4.2}, both climatological and post-processed forecasts (MLP and POLR models) result in far lower mean CRPS values than the raw EUPPBench ensemble for all lead times (see Figure \ref{fig:crps_crpssB}a); however, now the ranking of the competing predictions is different. As depicted in Figure \ref{fig:crps_crpssB}b, climatology performs surprisingly well, clearly outperformed for all forecast horizons only by the locally trained POLR and MLP approaches. Except for lead times corresponding to 1200 UTC, POLR-C and MLP-C also yield positive CRPSS with respect to climatology and are superior to MLP-L, whereas POLR-R and MLP-R follow a similar pattern but perform much worse than the corresponding semi-local models. The ranking of forecasts based on  Figure \ref{fig:crps_crpssB}b (POLR-L -- MLP-C -- POLR-C -- MLP-L -- Climatology -- MLP-R -- POLR-R) is also confirmed by Table \ref{tab3} reporting the overall improvement in mean scores with respect to the raw EUPPBench visibility ensemble forecasts.

\begin{figure}[t]
\begin{center}
\epsfig{file=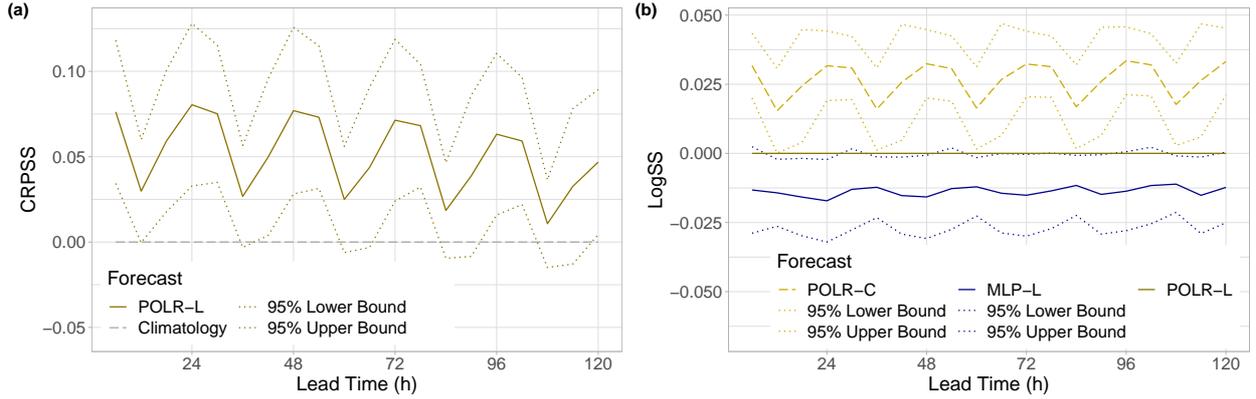, width=\textwidth}
\end{center}
\caption{CRPSS of POLR-L post-processed  EUPPBench visibility forecasts for calendar year 2018 with respect to climatology (a) and LogSS of POLR-C and MLP-L approaches with respect to the POLR-L model (b)  together with 95\,\% confidence intervals as functions of the lead time.}
\label{fig:crpss_logssB}
\end{figure}

In terms of the mean LogS, climatology is again far behind post-processed forecasts; Figure \ref{fig:logs_logssB}a is rather similar to Figure \ref{fig:logs_logss}a. According to Figure \ref{fig:logs_logssB}b, semi-local models clearly outperform their local counterparts and except for lead times corresponding to 1200 UTC, even the regional approaches are superior to the local ones. From the studied calibration methods POLR-C results in the lowest mean LogS, closely followed by the MLP-C, whereas MLP-L shows the poorest performance, see also Table \ref{tab3}.

\begin{figure}[h!]
\begin{center}
\epsfig{file=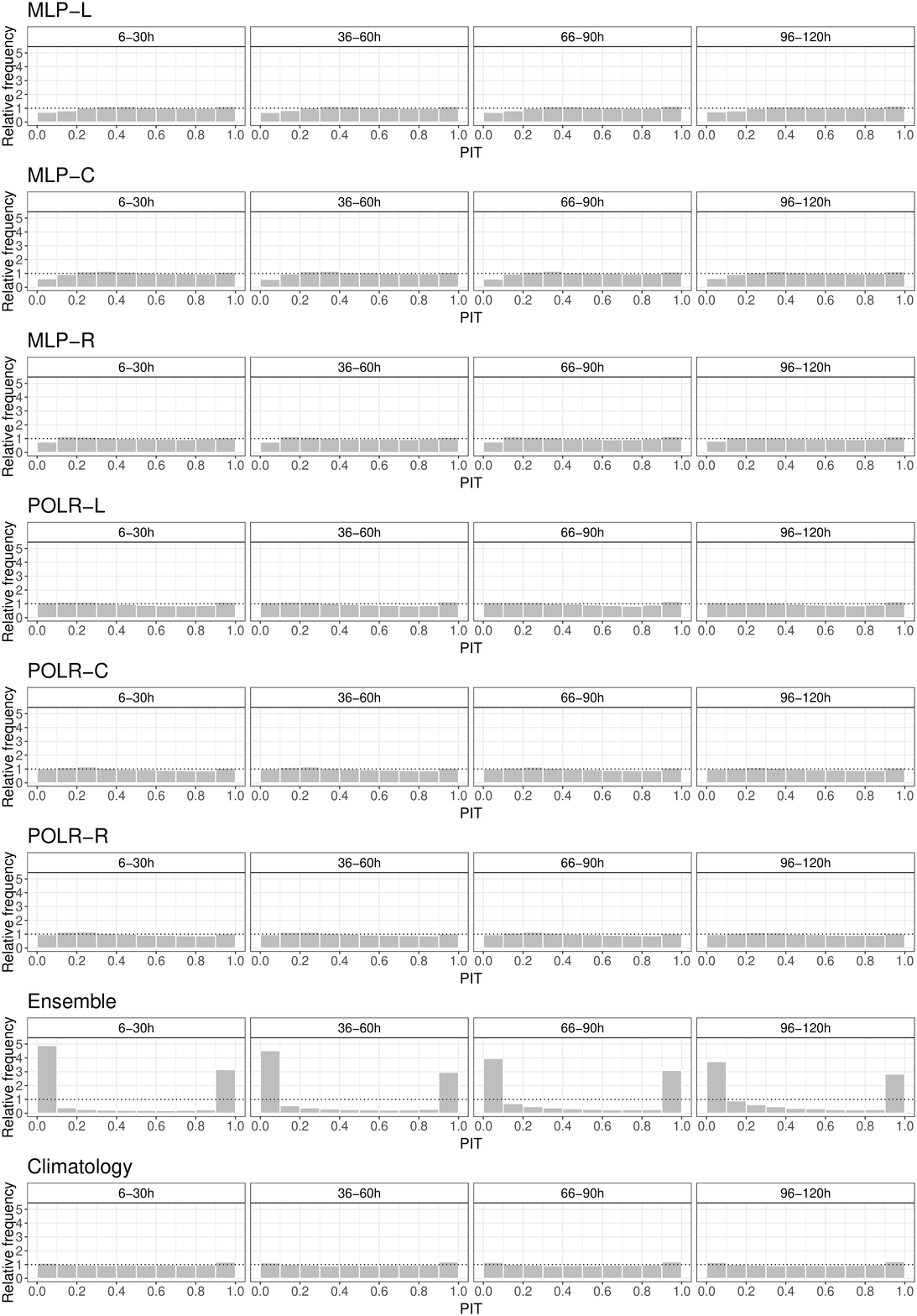, width=0.9\textwidth}
\end{center}
\caption{PIT histograms of post-processed, climatological and raw visibility EUPPBench forecasts for calendar year 2018 for lead times 6–30 h, 36–60 h, 66–90 h and 96–120 h.}
\label{fig:pitB}
\end{figure}

The good performance of climatology with regard to the mean CRPS is also confirmed by Figure \ref{fig:crpss_logssB}a, as at forecast horizons 36 h, 60 h, 66 h, 84 h, 90 h, 108 h and 114 h even the best performing POLR-L model fails to result in significantly positive CRPSS at a 5\,\% level. Furthermore, in contrast to the case study of Section \ref{subs4.2}, the LogSS of POLR-C with respect to POLR-L is significantly positive for all investigated lead times, whereas the 95\,\% upper bound for the LogSS of the least performing MLP-L almost coincides with the reference line. This means a barely significant difference between the MLP-L approach and POLR-L.

The PIT histograms of Figure \ref{fig:pitB} tell us the same story about calibration as the corresponding panels of Figure \ref{fig:pit}. Raw EUPPBench forecasts are highly underdispersive with a small bias increasing with the forecast horizon, whereas climatology and all post-processing approaches display rather flat histograms. The $\alpha^0_{1234}$ test again reject uniformity for all calibrated forecasts for all lead times; however, now the two regional approaches results in the highest overall mean $p$-values, see Table \ref{tab4}.

\begin{figure}[t]
\begin{center}
\epsfig{file=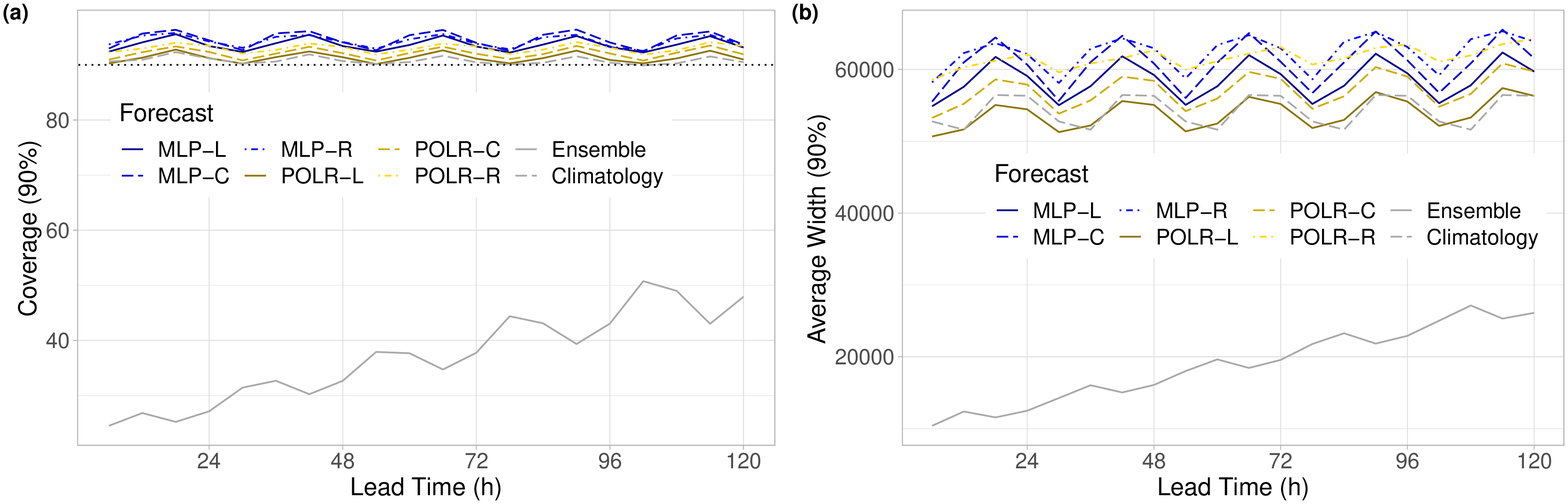, width=\textwidth}
\end{center}
\caption{Coverage (a) and average widths (b) of 90\,\% central prediction intervals of raw and post-processed EUPPBench visibility forecasts for calendar year 2018 as functions of the lead time. In panel (a) the ideal coverage is indicated by the horizontal dotted line.}
\label{fig:cov_awB}
\end{figure}

According to Figure \ref{fig:cov_awB}a, with the increase of the forecast horizon the coverage of 90\,\% central prediction intervals of the raw EUPPBench forecasts increases from 22\,\% to 51\,\%, and shows again a clear diurnal cycle.  This diurnal cycle is also preserved in the coverage values of climatological and post-processed forecasts; however, these curves do not display any further dependence on the forecast horizon and are just slightly above 90\,\%. Similar to Figure \ref{fig:cov_aw}a, the best coverage again corresponds to climatological forecasts with a mean absolute deviation from the nominal value of 0.82\,\%, followed by the POLR models (POLR-L: 1.32\,\%; POLR-C: 2.12\,\%; POLR-R: 3.03\,\%) and MLP approaches (MLP-L: 3.69\,\%; MLP-C: 4.59\,\%; MLP-R: 4.35\,\%). Again, the average widths of 90\,\% central prediction intervals depicted in Figure \ref{fig:cov_awB}b are nicely in line with the corresponding coverage values, with the POLR-L approach providing the sharpest post-processed forecasts.

\begin{figure}[t]
 \begin{center}
 \epsfig{file=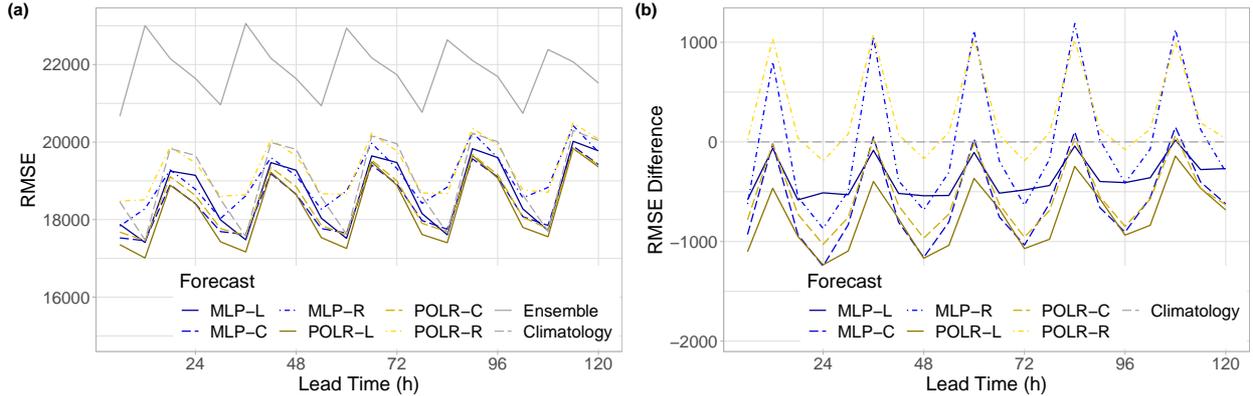, width=\textwidth}
 \end{center}
 \caption{RMSE of the mean EUPPBench visibility forecasts for calendar year 2018 (a) and difference in RMSE from climatology (b) as functions of the lead time.}
 \label{fig:rmse_rmsedB}
 \end{figure}

Furthermore, in line with the case study of Section \ref{subs4.2}, all post-processing methods result in a considerable decrease in the RMSE of the mean forecast (see Figure \ref{fig:rmse_rmsedB}a). Moreover, according to Figure \ref{fig:rmse_rmsedB}b, for the EUPPBench data climatology is fully able to catch up with the POLR and MLP models and for forecast horizons corresponding to 1200 UTC it definitely outperforms their local versions.

\begin{table}[t]
  \begin{center}
    \begin{tabular}{c|c|c|c|c|c}
MLP-L&MLP-C&MLP-R&POLR-L&POLR-C&POLR-R\\ \hline
$2.19\times 10^{-13}$&$1.80\times 10^{-14}$&$2.82\times 10^{-4}$&$5.55\times 10^{-18}$&$1.15\times 10^{-6}$&$4.57\times 10^{-4}$ \end{tabular}
\end{center}
\caption{Overall mean $p$-values of the $\alpha^0_{1234}$ tests for uniformity of the PIT values for calendar year 2018.}
\label{tab4}
\end{table}

Finally, 350-day climatology performs well again only in terms of the LogS; however, in this case it is still slightly behind the post-processed forecasts for all forecast horizons.

\section{Conclusions}
\label{sec5}

We investigate the predictive performance of the proportional odds logistic regression and multilayer perceptron neural network classifiers for statistical post-processing of visibility ensemble forecasts in the situation, when observations are reported in discrete values following the suggestions of the World Meteorological Organization. In two case studies, the forecast skill of the proposed calibration methods is tested on two different datasets of ECMWF visibility ensemble forecasts covering two overlapping geographical regions and two different time domains. First, we consider the 51-member operational forecasts for 30 locations in Central Europe with a forecast horizon of 240 h and a time step of 6 h for calendar years 2020 and 2021. The second dataset is part of the EUPPBench benchmark data, where the ECMWF high-resolution forecast is also included. It contains ensemble forecasts and validating observations for  42 stations in Germany and France for calendar years 2017 -- 2018 with a maximal lead time of 120 h and temporal resolution of 6 h.  Local, semi-local, and regional training is tested for both post-processing approaches, which are based on 350-day rolling training periods. As reference, we consider raw and climatological visibility forecasts where for the latter short rolling past time intervals are chosen in  order to capture seasonal variations in visibility observations. 

In general, climatological and post-processed forecasts outperform the raw ensemble in terms of the mean CRPS, the mean LogS, and the RMSE of the mean forecast over the verification data by a wide margin, but the difference decreases with the increase of the forecast horizon. They result in better coverage but wider central prediction intervals than the raw ECMWF visibility forecasts and the corresponding PIT histograms  are much closer to the uniform distribution than those of the raw ensemble. POLR models are superior to their MLP counterparts which is in line with the findings of \citet{bleab21} in the case of total cloud cover forecasts, and the performance of climatology is comparable with the skill of post-processed forecasts in all investigated scores but the mean LogS. From the competing post-processing methods the locally trained POLR model shows the best overall performance, whereas for the machine learning-based approaches the semi-local training is often superior to the local one.

The results of the current work open several avenues for future research. First, here as inputs of our post-processing models, we have used only features derived from the raw visibility forecasts. However, as recent studies demonstrate \citep[see e.g.][]{rl18,bleab21,sl22}, the incorporation of additional predictors, especially in the case of machine learning-based approaches, where the extension of the input feature set is rather straightforward, might significantly improve the forecast skill. Furthermore, so far we have focused on univariate forecasts for a single location and forecast horizon. However, in the last decade, a wide range of multivariate post-processing techniques have been developed, which are able to restore dependence structures lost during the univariate calibration, for an overview see e.g. \citet{lbm20} or \citet{llhb23}. The investigation of spatially and/or temporally consistent calibrated visibility forecasts might be an other interesting direction of our future work. Finally, as several SYNOP stations report visibility in 10 m steps which can be considered almost as continuous, using e.g. the interpolation technique of \citet{sswh20}, one might try to develop methods for estimating continuous predictive distributions for this weather quantity.

\bigskip
\noindent
{\bf Acknowledgments.} \  The authors gratefully acknowledge the support of the the \'UNKP-22-3 New National Excellence Program of The Ministry for Innovation and Technology and of the  National Research, Development and Innovation Office under Grant No. K142849. Finally, they are indebted to Zied Ben Bouall\`egue for providing the ECMWF visibility data.


\begin{thebibliography}{99}
\bibitem[Baran {\em et al.\/}, 2021]{bleab21}  Baran, \'A., Lerch, S., El Ayari,
  M. and Baran, S. (2021) Machine learning for total cloud cover prediction.
  {\em Neural. Comput. Appl.\/} {\bf 33}, 2605--2620.
  
\bibitem[Bauer {\em et al.\/}, 2015]{btb15} Bauer, P., Thorpe, A. and Brunet,
  G. (2015) The quiet revolution of numerical weather prediction. {\em
    Nature\/} {\bf 525}, 47--55.

\bibitem[Breiman, 2001]{breiman01} Breiman, L. (2001) Random forests. {\em
    Mach. Learn.\/} {\bf 45}, 5--32.  

\bibitem[Bremnes, 2019]{brem19} Bremnes, J. B. (2019) Constrained quantile
  regression splines for ensemble postprocessing. {\em Mon. Weather Rev.\/}
  {\bf 147}, 1769--1780.

\bibitem[Bremnes, 2020]{brem20} Bremnes, J. B. (2020) Ensemble postprocessing
  using quantile function regression based on neural networks and Bernstein
  polynomials. {\em Mon. Weather Rev.\/} {\bf 148}, 403--414.

\bibitem[Buizza, 2018a]{b18a} Buizza, R. (2018a) Introduction to the special
  issue on ``25 years of ensemble forecasting''. {\em Q. J. R. Meteorol.
    Soc.\/} {\bf 145}, 1--11.

\bibitem[Buizza, 2018b]{b18b} Buizza, R. (2018b) Ensemble forecasting and 
	the need for calibration. In Vannitsem, S., Wilks, D. S., Messner, J. W.
     (eds.), {\em Statistical Postprocessing of Ensemble Forecasts.\/}
     Elsevier, Amsterdam, pp. 15--48.

\bibitem[Buizza  {\em et al.\/}, 2005]{bhtp05} Buizza, R., Houtekamer, P. L.,
  Toth, Z., Pellerin, G., Wei, M. and Zhu, Y. (2005) A comparison of the ECMWF,
  MSC, and NCEP global ensemble prediction systems.  {\em Mon. Weather Rev.\/}
  {\bf 133}, 1076--1097.  
  
\bibitem[Chmielecki and Raftery, 2011]{cr11} Chmielecki, R. M. and Raftery, A.
  E. (2011) Probabilistic visibility forecasting using Bayesian model
  averaging. {\em Mon. Weather Rev.\/} {\bf 139}, 1626--1636.

\bibitem[Dabering {\em et al.\/}, 2017]{dmmz17} Dabernig, M., Mayr, G. J.,
  Messner, J. W. and Zeileis, A. (2017)  Spatial ensemble post-processing with
  standardized anomalies. {\em Q. J. R. Meteorol. Soc.\/} {\bf 143}, 909--916. 

\bibitem[Demaeyer {\em at al.\/}, 2023]{eupp} Demaeyer, J., Bhend, J., Lerch,
  S., Primo, C., Van Schaeybroeck, B., Atencia, A., Ben Bouall\`egue, Z.,
  Chen, J., Dabernig, M., Evans, G., Faganeli Pucer, J., Hooper, B., Horat, N.,
  Jobst, D., Mer\v{s}e, J., Mlakar, P., M\"oller, A., Mestre, O., Taillardat,
  M. and Vannitsem, S. (2023) The EUPPBench postprocessing benchmark dataset
  v1.0. {\em Earth Syst. Sci. Data\/}, doi:10.5194/essd-2022-465. 

\bibitem[ECMWF Directorate, 2012]{ecmwf12} ECMWF Directorate (2012) Describing
  ECMWF's forecasts and forecasting system. {\em ECMWF Newsletter\/} {\bf 133},
  11--13.  

\bibitem[ECMWF, 2021]{ifs21} 
 ECMWF 
  (2021) {\em IFS Documentation CY47R3 - Part IV Physical processes.\/} ECMWF,
  Reading. Available at: \url{http://dx.doi.org/10.21957/eyrpir4vj} [Accessed
  on 23 May 2023]

\bibitem[Friederichs and Hense, 2007]{fh07} Friederichs, P. and Hense, A.
  (2007) Statistical downscaling of extreme precipitation events using
  censored quantile regression. {\em Mon. Weather Rev.\/} {\bf 135},
  2365--2378.
  
\bibitem[Friedman, 2001]{fried01} Friedman, J. H. (2001) Greedy function
  approximation: a gradient boosting machine. {\em Ann. Stat.\/} {\bf 29},
  1189--1232.

\bibitem[Ghazvinian {\em et al.\/}, 2021]{gzshf21} Ghazvinian, M., Zhang, Y.,
  Seo, D-J., He, M. and Fernando, N. (2021) A novel hybrid artificial neural
  network - parametric scheme for postprocessing medium-range precipitation
  forecasts.  {\em Adv. Water Resour.\/} {\bf 151}, paper 103907.

\bibitem[Gneiting {\em et al.\/}, 2007]{gbr07} Gneiting, T.,
  Balabdaoui, F. and Raftery, A. E. (2007) Probabilistic forecasts,
  calibration and sharpness. {\em J. R. Stat. Soc. Series B Stat. Methodol.\/}
  {\bf 69}, 243--268.    
  
\bibitem[Gneiting and Raftery, 2005]{gr05} Gneiting, T. and Raftery, A. E.
   (2005) Weather forecasting with ensemble methods. {\em Science\/} {\bf 310},
   248--249.

\bibitem[Gneiting and Raftery, 2007]{gr07} Gneiting, T. and Raftery, A. E.
  (2007) Strictly proper scoring rules, prediction and estimation. {\em J.
    Amer. Statist. Assoc.\/} {\bf 102}, 359--378.     

\bibitem[Gneiting {\em et al.\/}, 2005]{grwg05} Gneiting, T., Raftery, A. E.,
  Westveld, A. H. and Goldman, T. (2005) Calibrated probabilistic forecasting
  using ensemble model output statistics and minimum CRPS estimation. {\em Mon.
    Weather Rev.\/} {\bf 133}, 1098--1118.

\bibitem[Gneiting and Ranjan, 2013]{gr13} Gneiting, T. and Ranjan, R. (2013)
  Combining predictive distributions. {\em  Electron. J. Stat.\/} {\bf 7},
  1747--1782.

\bibitem[Good, 1952]{good52} Good, I. J. (1952) Rational decisions.
  {\em J. R. Stat. Soc. Series B Stat. Methodol.\/} {\bf 14}, 107--114.    

\bibitem[Goodfellow {\em et al.\/}, 2016]{dlbook} Goodfellow, I., Bengio, Y.
  and Courville, A. (2016) {\em Deep Learning.\/} MIT Press, Cambridge.   

\bibitem[Haiden {\em et al.\/}, 2021]{ecmwfEval21} Haiden, T., Janousek, M.,
   Vitart, F.,  Ben Bouall\`egue, Z., Ferranti, L., Prates, F. and Richardson,
   D. (2021) Evaluation of ECMWF forecasts, including the 2021 upgrade.
   {\em ECMWF Technical Memorandum\/} No. 884. Available at:
   \url{http://dx.doi.org/10.21957/90pgicjk4} [Accessed on 23 May 2023]

\bibitem[Hemri {\em et al.\/}, 2016]{hhp16} Hemri, S., Haiden, T. and
  Pappenberger, F. (2016) Discrete postprocessing of total cloud cover ensemble
  forecasts. {\em Mon. Weather Rev.\/} {\bf 144}, 2565--2577.

\bibitem[Hemri {\em et al.\/}, 2014]{hspbh14} Hemri, S., Scheuerer, M., 
  Pappenberger, F., Bogner, K. and Haiden, T. (2014) Trends in the predictive 
  performance of raw ensemble weather forecasts. {\em Geophys. Res. Lett.\/} 
  {\bf 41}, 9197--9205.  

\bibitem[Izenman, 2008]{izen08} Izenman, A. J. (2008) {\em Modern Multivariate
    Statistical Techniques. Regression, Classification and Manifold
    Learning.\/} Springer, New York.

\bibitem[Kn\"uppel, 2015]{k15} Kn\"uppel, M. (2015)
  Evaluating the calibration of multi-step-ahead density forecasts using raw
  moments. {\em J. Bus. Econ. Stat.\/} {\bf 33}, 270--281.

\bibitem[Lakatos {\em et al.\/}, 2023]{llhb23} Lakatos, M., Lerch, S., Hemri,
  S. and Baran, S. (2023) Comparison of multivariate post-processing methods
  using global ECMWF ensemble forecasts. {\em Q. J. R. Meteorol. Soc.\/}
  {\bf 149}, 856--877.    
  
\bibitem[Lerch and Baran, 2017]{lb17} Lerch, S. and Baran, S. (2017)
  Similarity-based semi-local estimation of EMOS models. {\em J. R. Stat. Soc.
    Ser. C Appl. Stat.\/} {\bf 66}, 29--51.  

\bibitem[Lerch {\em et al.\/}, 2020]{lbm20} Lerch, S., Baran, S., M\"oller, A.,
  Gro\ss, J., Schefzik, R., Hemri, S. and Graeter, M. (2020) Simulation-based
  comparison of multivariate ensemble post-processing methods. {\em Nonlinear
    Process. Geophys.\/}  {\bf 27}, 349--371.     

\bibitem[McCullagh, 1980]{mccull80} McCullagh, P. (1980) Regression model for
  ordinal data (with discussion). {\em J. R. Stat. Soc. Series B Stat.
    Methodol.\/} {\bf 42}, 243--268.  
  
\bibitem[Molteni {\em et al.\/}, 1996]{mbpp96} Molteni, F., Buizza, R.,
  Palmer, T. N. and Petroliagis, T. (1996) The ECMWF ensemble prediction
  system: Methodology and validation. {\em Q. J. R. Meteorol. Soc.\/} {\bf
    122}, 73--119.

\bibitem[Murphy, 1973]{murphy73} Murphy, A. H. (1973) Hedging and skill
  scores for probability forecasts. {\em J. Appl. Meteorol.\/} {\bf 12},
  215--223.

\bibitem[Politis and Romano, 1994]{pr94} Politis, D. N. and Romano, J. P. (1994)
  The stationary bootstrap. {\em J. Amer. Statist. Assoc.\/} {\bf 89},
  1303--1313.    

\bibitem[Raftery {\em et al.\/}, 2005]{rgbp05} Raftery, A. E., Gneiting, T.,
  Balabdaoui, F. and Polakowski, M. (2005) Using Bayesian model averaging to
  calibrate forecast ensembles. {\em Mon. Weather Rev.\/} {\bf 133},
  1155--1174.

\bibitem[Rasp and Lerch, 2018]{rl18} Rasp, S. and Lerch, S. (2018) Neural
  networks for postprocessing ensemble weather forecasts. {\em Mon. Weather
    Rev.\/} {\bf 146}, 3885--3900.

\bibitem[Scheuerer {\em et al.\/}, 2020]{sswh20}  Scheuerer, M., Switanek, M.
  B., Worsnop, R. P. and Hamill, T. M. (2020) Using artificial neural networks
  for generating probabilistic subseasonal precipitation forecasts over
  California. {\em Mon. Weather Rev.\/} {\bf 148}, 3489--3506.

\bibitem[Schultz and Lerch, 2022]{sl22}  Schultz, B. and Lerch, S. (2022)
  Machine learning methods for postprocessing ensemble forecasts of wind gusts:
  a systematic comparison. {\em Mon. Weather Rev.\/} {\bf 150}, 235--257.

\bibitem[Taillardat {\em et al.\/}, 2016]{tmzn16} Taillardat, M., Mestre, O.,
  Zamo, M. and Naveau, P. (2016) Calibrated ensemble forecasts using quantile
  regression forests and ensemble model output statistics. {\em Mon. Weather
    Rev.\/} {\bf 144}, 2375--2393.

\bibitem[Thorarinsdottir and Gneiting, 2010]{tg10}  Thorarinsdottir, T. L. and
  Gneiting, T. (2010) Probabilistic forecasts of wind speed: ensemble model
  output statistics by using heteroscedastic censored regression. {\em J. R.
    Stat. Soc. Ser. A Stat. Soc.\/} {\bf 173},  371--388.  

\bibitem[Vannitsem {\em et al.\/}, 2021]{vbd21}Vannitsem, S., Bremnes, J. B.,
  Demaeyer, J., Evans, G. R., Flowerdew, J., Hemri, S., Lerch, S., Roberts, N.,
  Theis, S., Atencia, A., Ben Boual\`egue, Z., Bhend, J., Dabernig, M., De
  Cruz, L., Hieta, L., Mestre, O., Moret, L., Odak Plenkovi\v{c}, I., Schmeits,
  M., Taillardat, M., Van den Bergh, J., Van Schaeybroeck, B., Whan, K. and
  Ylhaisi, J. (2021) Statistical postprocessing for weather forecasts --
  review, challenges and avenues in a big data world. {\em Bull. Amer.
    Meteorol. Soc.\/} {\bf 102}, E681--E699.

\bibitem[Venables and Ripley, 2002]{vr02} Venables, W. N. and Ripley, B. D.
  (2002) {\em Modern Applied Statistics with S. Fourth Edition.\/} Springer,
  New York.     

\bibitem[Wilks, 2018]{w18} Wilks, D. S. (2018) Univariate ensemble
  postprocessing. In Vannitsem, S., Wilks, D. S., Messner, J. W. (eds.), {\em
    Statistical Postprocessing of Ensemble Forecasts.\/} Elsevier, Amsterdam,
  pp. 49--89.

\bibitem[Wilks, 2019]{w19} Wilks, D. S. (2019) {\em Statistical Methods in the
    Atmospheric Sciences. 4th ed.\/} Elsevier, Amsterdam.    
  
\bibitem[WMO, 1992]{wmo92} World Meteorological Organization (1992) 
    {\em International Meteorological Vocabulary\/} (WMO-No.182). WMO, Geneva.

\bibitem[WMO, 2018]{wmo18} World Meteorological Organization (2018) {\em Guide
    to Instruments and Methods of Observation. Volume I – Measurement of
    Meteorological Variables\/} (WMO-No.8). WMO, Geneva.

\bibitem[Zell {\em et al.\/}, 1994]{snns94} Zell, A., Mache, N., H\"ubner, R.,
  Mamier, G., Vogt, M., Schmalzl, M. and  Herrmann, K.-U. (1994) SNNS
  (Stuttgart Neural Network Simulator). In Skrzypek, J. (ed.), {\em
    Neural Network Simulation Environments.\/} The Kluwer International Series
  in Engineering and Computer Science, vol 254. Springer, Boston.

\bibitem[Zhou {\em et al.\/}, 2009]{zdmqd09} Zhou, B., Du, J., McQueen, J. and
  Dimego, G. (2009) Ensemble forecast of ceiling, visibility, and fog with
  NCEP Short-Range Ensemble Forecast system (SREF). {\em Aviation, Range,
    and Aerospace Meteorology Special Symposium on Weather–Air Traffic
    Management Integration\/}, Phoenix, AZ, American Meteorological Society,
  extended abstract 4.5. Available at: \url{https://ams.confex.com/ams/89annual/techprogram/paper_142255.htm}
  [Accessed on 23 May 2023]

\bibitem[Zhou {\em et al.\/}, 2012]{zdgd12}  Zhou, B., Du, J., Gultepe, I.
  and Dimego, G. (2012) Forecast of low visibility and fog from NCEP: Current
  status and efforts. {\em Pure Appl. Geophys.\/} {\bf 169}, 895--909.

\end{thebibliography}
\end{document}